\documentstyle[aps,prl,amssymb,epsfig]{revtex}
\begin{document}
\draft
\twocolumn[\hsize\textwidth\columnwidth\hsize\csname %
@twocolumnfalse\endcsname

\title{Electron momentum distribution in underdoped cuprates}
\author{A. Ram\v sak$^{1,2}$, I. Sega$^1$, and P. Prelov\v sek$^{1,2}$}
\address{ $^{1}$ J. Stefan Institute, 1000 Ljubljana, Slovenia }
\address{$^{2}$ Faculty of Mathematics and Physics, University of
Ljubljana, 1000 Ljubljana, Slovenia }
\date{\today}
\maketitle         
\begin{abstract}\widetext
We investigate the electron momentum distribution function (EMD) in a
weakly doped two-dimensional quantum antiferromagnet (AFM) as
described by the $t$-$J$ model.  
Our analytical results for a single
hole in an AFM based on the self-consistent Born approximation (SCBA)
indicate an anomalous momentum dependence of EMD
showing 'hole pockets' coexisting with a signature of an emerging
large Fermi surface. The position of the incipient Fermi surface and
the structure of the EMD is determined by the momentum of the ground
state. Our analysis shows that this result remains robust in the
presence of next-nearest neighbor hopping terms in the model.
Exact diagonalization results for small clusters are with the SCBA
reproduced quantitatively.
\end{abstract}
 ]
\narrowtext
One of the most intriguing questions concerning the superconducting
cuprates is the existence and the character of the Fermi surface (FS),
in particular in their underdoped regime.  
This problem has been intensively studied 
experimentally with the angle-resolved
photoemission spectroscopy (ARPES)
\cite{shen95,wells95,marshall96,norman98}. 
There have been also several theoretical investigations of this
problem, using the exact diagonalization (ED) of small clusters
\cite{stephan91,eder98,chernyshev98}, string calculations
\cite{eder91}, slave-boson theory \cite{lee96} and the high
temperature expansion \cite{putikka98}.  While a consensus has been
reached about the existence of a large Fermi surface in the
optimum-doped and overdoped materials, in the interpretation of ARPES
experiments for the {\it underdoped} cuprates
\cite{marshall96} the issue of the debate is (i) why are
experiments more consistent with the existence of parts of large FS,
i.e., rather Fermi arcs or Fermi patches \cite{norman98,furukawa98}
than with a 'hole pocket' type small FS, predicted by several
theoretical methods based on the existence of AFM long range order in
cuprates, (ii) how does a partial FS eventually evolve with doping
into a large closed one.

The electron momentum distribution function $n_{\bf k}=\langle
\Psi_{{\bf k}_0}|\sum_\sigma c_{{\bf k},\sigma}^\dagger c_{{\bf
k},\sigma} |\Psi_{{\bf k}_0} \rangle$ is the key quantity for
resolving the problem of the Fermi surface.  In this paper we study
the EMD for $|\Psi_{{\bf k}_0} \rangle$ which represents a weakly
doped AFM, i.e, it is the ground state (GS) wave function of a planar
AFM with one hole and the GS wave vector ${\bf k}_0$. In the present
work we investigate the low-energy physics of the CuO$_2$ planes in
cuprates within the framework of the standard $t$-$J$ model with
nearest-neighbor hopping $t_{ii'}\equiv t$ and the AFM exchange $J$.
In order to come closer to 
the realistic situation in cuprates the
model is extended with the next-nearest-neighbor hopping
$t_{ii^\prime}\equiv t^\prime$ and the third-neighbor hopping terms
$t_{ii^\prime} \equiv t^{\prime \prime}$, for $ii^\prime$ representing
next-nearest-neighbors and third-neighbors, respectively\cite{kyung96} ,
\begin{eqnarray}
H&=& -\sum_{<ii^\prime>\sigma} t_{ii^\prime} \bigl(
{\tilde c}_{i,\sigma}^\dagger
{\tilde c}_{i^\prime,\sigma} + \mbox{H.c.} \bigr)+ \nonumber \\
& &+ J \sum_{<ij>}
\bigl[ S_i^z S_j^z+\frac{\gamma}{2}(S_i^+ S_j^- + S_i^- S_j^+ ) \bigr].
\label{tj}
\end{eqnarray}
${\tilde c}_{i,\sigma}^\dagger$ (${\tilde c}_{i,\sigma}$) are
electron creation (annihilation) operators acting in a space
forbidding double occupancy on the same site. 
The effect of double occupancy \cite{eskes96}
on $n_{\bf k}$ is not studied in the present framework of the $t$-$J$ model.
$S^\alpha_i$ are spin
operators. For convenience we treat the anisotropy $\gamma$ as a free
parameter, with $\gamma=0$ in the Ising limit, and $\gamma\to1$
in the Heisenberg model. Recent studies of the $t$-$J$ model
with $t'$, $t''$ terms included have shown a very good agreement of the
calculated quasiparticle (QP) dispersion with experimental results of
ARPES \cite{kyung96} whereby quantitative differences between
different Cu compounds have been attributed to different values of
$t^\prime$ and $t^{\prime \prime}$ \cite{feiner96}.

Our analytical approach is based on a spinless fermion -- Schwinger boson
representation of the $t$-$J$ Hamiltonian \cite{schmitt88} and on the
SCBA for calculating both the Green's function
\cite{schmitt88,ramsak90,martinez91} and the corresponding wave function
\cite{reiter94,ramsak93}. The method is known to be successful in
determining spectral and other properties of the 
QP. In contrast to other methods the SCBA is expected to describe the 
{\it long-wavelength} physics since it is determined by the linear
dispersion of spin waves. The
{\it short-wavelength} properties 
can be studied with various methods, here we compare
the SCBA results with the corresponding ED, as shown further-on.

In the SCBA fermion operators are decoupled into
hole and pseudo spin - local boson operators: ${\tilde
c}_{i,\uparrow}\!=\!h^\dagger_i$, ${\tilde
c}_{i,\downarrow}\!=\!h^\dagger_i S^+_i \!\sim\! h^\dagger_i a_i$ and
${\tilde c}_{i,\downarrow}\!=\!h^\dagger_i$, ${\tilde
c}_{i,\uparrow}\!=\!h^\dagger_i S^-_i \!\sim\! h^\dagger_i a_i$ for
$i$ belonging to $A$- and $B$-sublattice, respectively.  The effective
Hamiltonian emerges
\cite{schmitt88,ramsak90,martinez91,bala95}
\begin{eqnarray}
{\tilde H}&=&
\sum_{\bf k} \epsilon^0_{\bf k} h^\dagger_{\bf k} h_{\bf k}
+\sum_{\bf q} \omega_{\bf q} \alpha^\dagger_{\bf q} \alpha_{\bf q}+
\nonumber\\
& &+N^{-1/2}\sum_{{{\bf k}}{{\bf q}}} (
M_{{\bf k}{\bf q}}
h_{{{\bf k}}-{{\bf q}} }^\dagger  h_{{{\bf k}} }
\alpha_{{\bf q}}^\dagger+{\rm H.c.} ),\label{lsw}
\end{eqnarray}
where $h_{\bf k}^\dagger$ is the creation operator for a (spinless)
hole in a Bloch state with a dispersion $\epsilon^0_{\bf
k}=4t^\prime\cos k_x
\cos k_y+2t^{\prime \prime}(\cos 2k_x- \cos 2k_y)$. The
AFM boson operator $\alpha^\dagger_{\bf q}$ creates an AFM magnon with
the energy $\omega_{\bf q}$, $M_{{\bf k}{\bf q}}$ is the
fermion-magnon coupling and $N$ is the number of lattice sites.

We calculate the Green's function for a hole $G_{\bf k}(\omega)$
within the SCBA \cite{schmitt88,ramsak90,martinez91}.
This approximation amounts to the summation of non-crossing diagrams
to all orders and the corresponding ground state wave function with
momentum ${\bf k}_0$ and energy $\epsilon_{{\bf k}_0}$
\cite{reiter94,ramsak93} is represented as
\begin{eqnarray}
|\Psi_{{\bf k}_0}\rangle&=&Z^{1/2}_{{\bf k}_0} \Bigl[
h_{{\bf k}}^\dagger
+N^{-1/2}\sum_{{\bf q}_1} M_{{\bf k}_0{\bf q}_1}
G_{\bar{\bf k}_1}(\bar\omega_1)
h_{\bar{\bf k}_1}^\dagger \alpha_{{\bf q}_1}^\dagger+  \nonumber\\
& &...+N^{-n/2}\!\!\!\sum_{{\bf q}_1,...,{\bf q}_n}
\!\!\!
M_{{\bf k}{\bf q}_1}
G_{\bar{\bf k}_1}(\bar\omega_1) \,...\,
M_{\bar{\bf k}_{n-1}{\bf q}_n} \times \nonumber\\
& &\quad\quad \times\;
G_{\bar{\bf k}_n}(\bar\omega_n)
\;h_{\bar{\bf k}_n}^\dagger
\alpha_{{\bf q}_1}^\dagger...\,\alpha_{{\bf q}_n}^\dagger  +\,\,\,
... \,\,\,
\Bigr]
|0\rangle. \label{psi}
\end{eqnarray}
Here $\bar{\bf k}_m={\bf k}_0\!-\!{\bf q}_1\!-\!...\!-\!{\bf q}_m$,
$\bar\omega_m=\epsilon_{{\bf k}_0}\!-\!\omega_{{\bf q}_1}\!-\!...\!-\!
\omega_{{\bf q}_m}$ and
$Z_{{\bf k}_0}$ is the QP spectral weight.
The wave function is properly normalized
$\langle\Psi_{{\bf k}_0}|\Psi_{{\bf k}_0}\rangle=1$
provided the number of magnon terms $n \to \infty$  \cite{ramsak93}.

The wave function Eq.~(\ref{psi}) corresponds to the projected space of
the model Eq.~(\ref{tj}) and therefore  the EMD is
$n_{\bf k}=\langle \Psi_{{\bf k}_0}|
{n_{\bf k}}|\Psi_{{\bf k}_0} \rangle
=\langle \Psi_{{\bf k}_0}|
{\tilde n_{\bf k}}|\Psi_{{\bf k}_0} \rangle$ with the
projected {\it electron} number operator ${\tilde n_{\bf k}}=\sum_\sigma
{\tilde c}_{{\bf k},\sigma}^\dagger  {\tilde c}_{{\bf k},\sigma}$.
Consistent with the SCBA approach, we decouple the latter
into hole and magnon operators,
\begin{eqnarray}
{\tilde n_{\bf k}}&=&\frac1N {\sum_{i j}} e^{-{\bf k} \cdot
({\bf R}_{i}-{\bf R}_{j})} h_i  h_j^\dagger\times \nonumber\\
& &\times \biggl(\eta^+_{ij} \bigl[1+a_i^\dagger a_j (1-\delta_{ij})
\bigr]
+ \eta^-_{ij}(a_i^\dagger + a_j)\biggr),\label{nk}
\end{eqnarray}
where $\eta^\pm_{ij}=(1\pm e^{-{\bf Q} \cdot ({\bf R}_{i}-{\bf
R}_{j})})/2$ with ${\bf Q}=(\pi,\pi)$.  Local $a^\dagger_i$ are
further expressed with proper magnon operators $\alpha^\dagger_{\bf
q}$. It should be noted that $n_{\bf k}$ should obey the sum rule
${\bar n}=\frac1N\sum_{\bf k} n_{\bf k}=1-c_{\rm h}$ and the
constraint $n_{\bf k} \leq 1+c_{\rm h}$, where $c_{\rm h}$ is the
concentration of holes\cite{stephan91,jaklic97}.

In general the expectation value $n_{\bf k}$ for a single hole
has to be calculated numerically and has the following structure
\begin{equation}
n_{\bf k}=1-\frac12 Z_{{\bf k}_0}(\delta_{{\bf k} {\bf k}_0}+
\delta_{{\bf k} {\bf k}_0+{\bf Q}})+{1 \over N}
\delta n_{\bf k}.\label{dnk}
\end{equation}
Here the second term proportional to $\delta$-functions corresponds
to 'hole pockets'. Note that $\delta n_{\bf
k}$, for the case of a single hole fulfills the sum
rule $\frac1N\sum_{\bf k} \delta n_{\bf k}=Z_{{\bf k}_0}-1$ and
$\delta n_{\bf k}\le1$. The introduction of $\delta n_{\bf k}$ is
convenient as it allows the comparison of results obtained with
different methods and on clusters of different size $N$.

For the case of Ising limit, $\gamma=0$, the Green's
function in the SCBA is independent of ${\bf k}$, $G_{\bf
k}(\omega)=G_0(\omega)$.  Therefore it is possible to express all
required matrix elements of $n_{\bf k}$ analytically and to perform a
summation of corresponding non-crossing contributions to any order
$n$, similar to Ref.~\cite{ramsak93}.  The result for $J/t=0.3$
(as relevant to cuprates) is
for some selected directions in the BZ presented  in Fig.~1.
We have also checked the convergence of $\delta n_{\bf
k}$ with the number of magnon lines, $n$.  For $J/t \agt 0.3$ we find
for all ${\bf k}$ that the contribution of terms $n>3$ amounts to less
than few percent. This is in agreement with the convergence of the
norm of the wave function, which is even faster \cite{ramsak93}. 
In Fig.~2(a) we present this $\delta
n_{\bf k}$ for the whole BZ. Here we note one interesting feature in
the Ising limit, i.e., the dip of $\delta n_{\bf k}$ in the center of
the BZ at $k \sim 0$ in agreement with Ref.~\cite{eder91}.
\vskip -.3 cm
\begin{figure}[htb]
\begin{center}
\epsfig{file=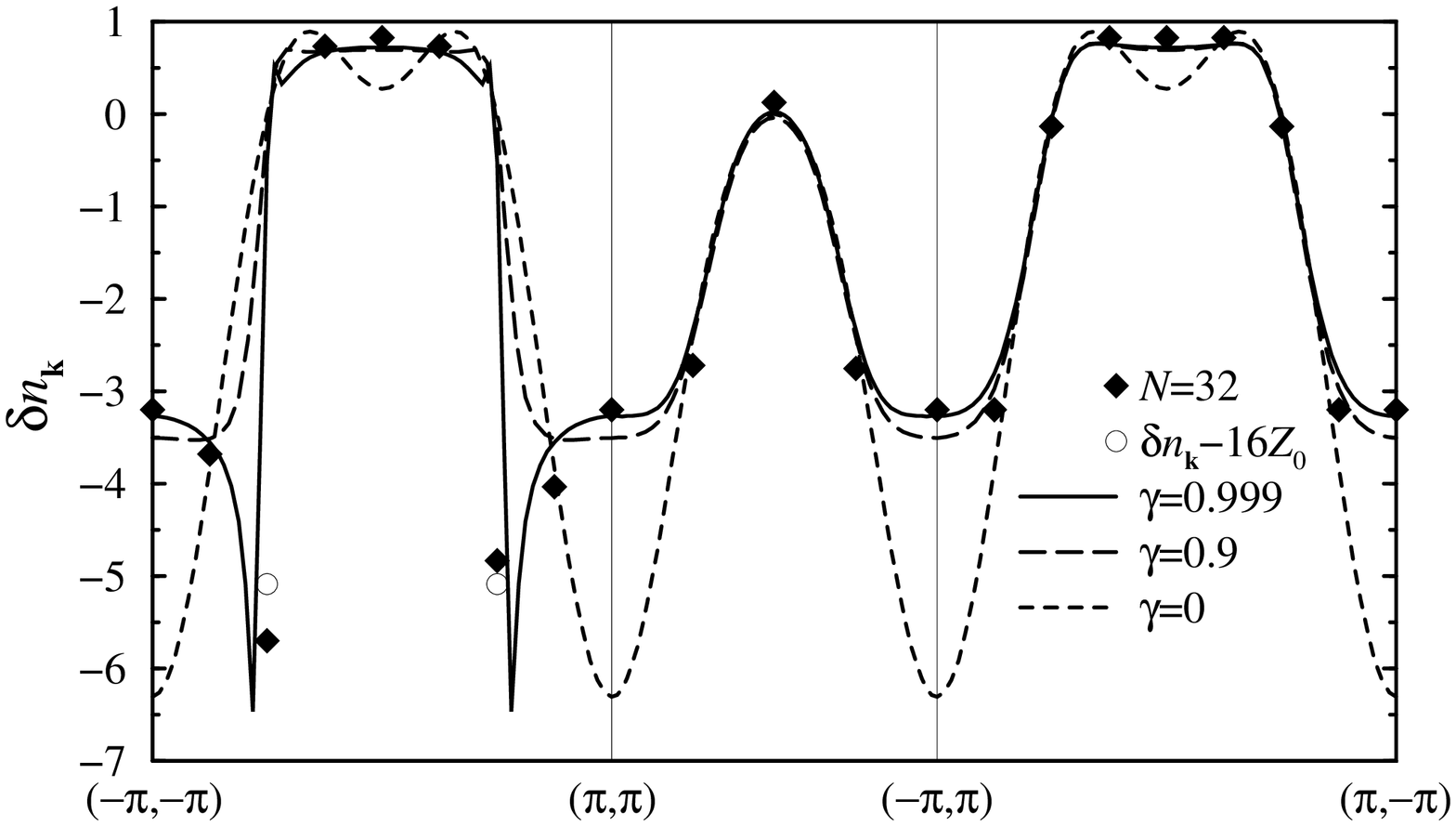,height=85mm,angle=-90}
\epsfig{file=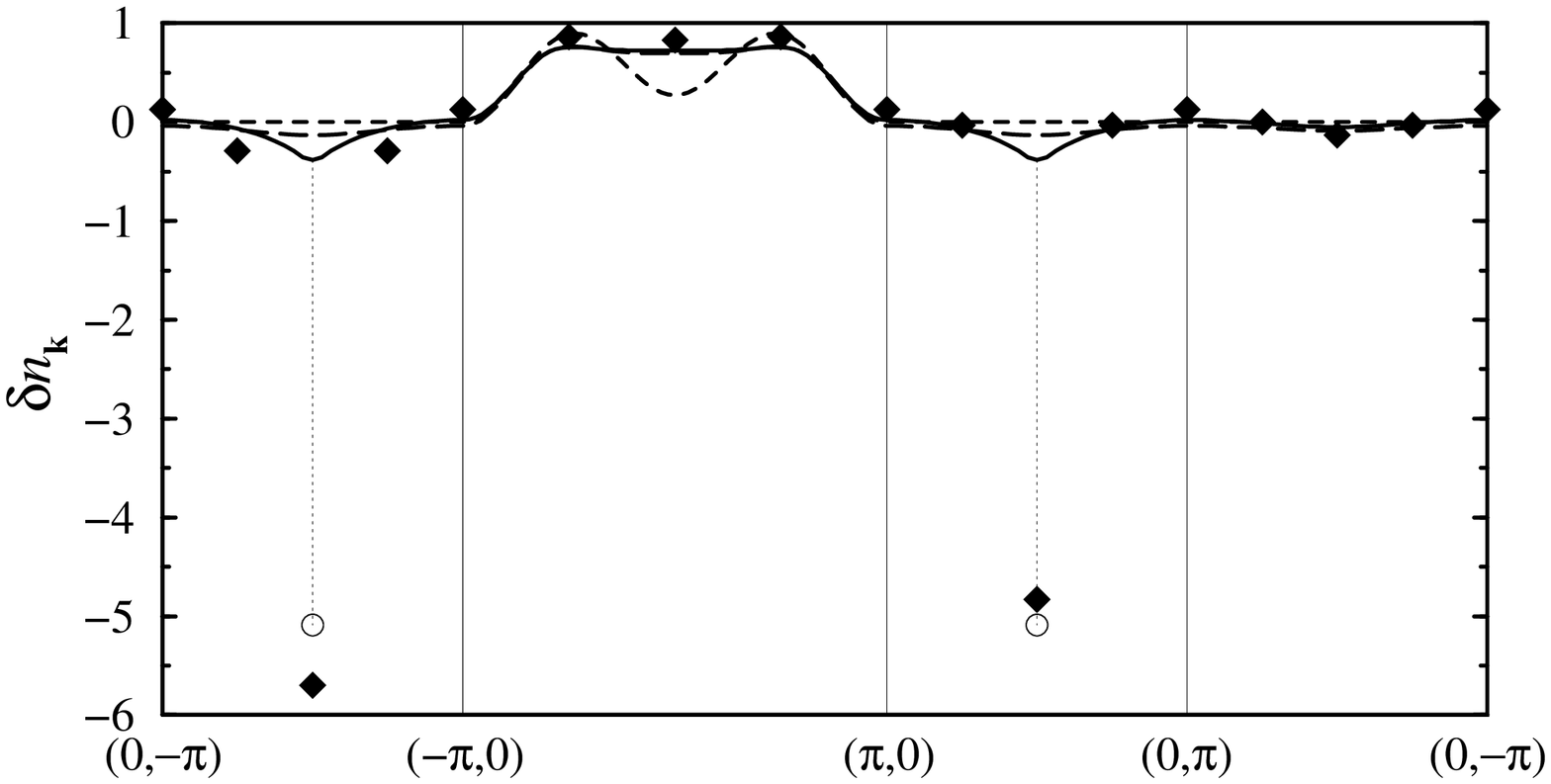,height=85mm,angle=-90}
\end{center}
\caption{$\delta n_{\bf k}$ along selected paths in the BZ for
$\gamma=0$, $0.9$, $0.999$. Full diamonds
represent ED results $N_{\rm }(n_{\bf k}-1)$ for $N_{\rm }=32$ 
of Chernyshev {\it et al.} \protect\cite{chernyshev98}.
Open circles represent the SCBA result for $\delta n_{{\bf
k}_0}-\frac12 N_{\rm } Z_{{\bf k}_0}$.  }
\end{figure}

Now we turn to the Heisenberg model, $\gamma \to 1$. Here the
important ingredient is the gap-less magnons
with linear dispersion and a more complex ground state of the planar
AFM.  $G_{\bf k}(\omega)$ and $\epsilon_{\bf k}$ become strongly ${\bf
k}$-dependent.  As a consequence $n_{\bf k}$ is now in general
dependent both on ${\bf k}$ and ${\bf k}_0$. The ground
state is for the $t$-$J$ model fourfold degenerate and we choose ${\bf
k}_0=(\frac\pi2,\frac\pi2)$. Results should be averaged over
all four possible ground state momenta if compared with, e.g., high
temperature expansion results \cite{putikka98} or discussed in connection with
ARPES data.

Let us first discuss the result in the limit $t/J\to0$, i.e., for a
{\it static} hole. In the linearized model, Eq.~(\ref{lsw}), $M_{\bf k
q}=0$. Therefore the hole is {\it not} coupled to the AFM ($Z_{{\bf k}_0}
\equiv 1$) and $\delta n_{\bf k}$ should be zero. However, a
straightforward use of Eq.~(\ref{nk}) leads also to non-vanishing and
momentum dependent
$\delta n_{\bf k}= n_0({\bf k})$. We attribute this momentum dependence
to the improper decomposition of ${\tilde c_{i,\sigma}}$ into
linearized pseudo spin operators instead of Schwinger bosons obeying
the local constraints \cite{schmitt88}. In the results for $\delta
n_{\bf k}$ presented in this paper (also at finite $J/t$) the
contribution $n_0({\bf k})$ is not included.

In Fig.~1 we present $\delta n_{\bf k}$ for the $t$-$J$ model 
with (almost) isotropic Heisenberg exchange, $\gamma=0.999$.
Numerical calculations are henceforth
performed for $J/t=0.3$ and $N$ corresponding to a $64 \times 64$
sites cluster \cite{ramsak90}. In evaluating the matrix elements we
take into account only terms with up to $n=3$ magnon lines in $
|\Psi_{{\bf k}_0}\rangle $ \cite{ramsak93}. The sum rule of our
numerical results for $\delta n_{\bf k}$ is nevertheless fulfilled
$\agt 96 \%$. 

Also included in Fig.~1 are results obtained with ED of the $N_{\rm }=32$
sites cluster\cite{chernyshev98}. These data are scaled 
as the quantity $N_{\rm }(n_{\bf k}-1)$ which should be directly compared
with the SCBA result $\delta n_{\bf k}$. At momenta ${\bf
k}={\bf k}_0$, ${\bf k}_0 +{\bf Q}$, however, one has to take into account
contributions from 'hole pocket' terms proportional to $\delta_{{\bf
k} {\bf k}_0}$ and with the scaling $\propto N_{\rm }$.  
Thus ED data should be compared at 
{\it these points} with $\delta n_{{\bf
k}_0}-\frac12 N_{\rm }Z_{{\bf k}_0}$ calculated from the 
SCBA. Note also, that the SCBA result for $\gamma=0.9$ (also
presented in Fig.~1) represents an intermediate step between the Ising
limit and the Heisenberg limit: the dip at the $\Gamma$ point, which
is in the Ising limit well pronounced here disappears
but the difference for directions ${\bf k}\parallel{\bf k}_0$ and ${\bf
k}\perp {\bf k}_0$ is not yet developed.
In Fig.~2(b) we present $\delta n_{\bf k}$ for the Heisenberg limit
($\gamma=0.99)$ in the entire BZ. In comparison with the Ising limit,
Fig.~2(a), $\delta n_{\bf k}$ exhibits a very strong momentum
dependence around $\pm{\bf k}_0$.

The comparison of the SCBA with ED results shows a quantitative
agreement at all points in the BZ. However, the SCBA result is {\it
symmetric} around $\Gamma$ point in the direction ${\bf k}\parallel
{\bf k}_0$, while small system results show a weak asymmetry for ${\bf
k}=\pm {\bf k}_0$, respectively. From our analysis of the
SCBA results for $N\to \infty$ and long range AFM spin background it
follows that $n_{\bf k}$ is in the thermodynamic limit $c_{\rm
h}\to0$ symmetric.  The asymmetry is in Ref.~\cite{chernyshev98}
attributed to the opening of the gap in the magnon spectrum at ${\bf
q}\sim{\bf Q}$ in finite systems. Within the SCBA the asymmetry also
appears if the EMD is evaluated with ${\bf k}_0$ {\it displaced} from
$(\frac\pi2,\frac\pi2)$ by a small amount $\delta{\bf k}_0$ (not shown
here). A common feature of finite clusters is a non-vanishing expectation
value of the current operator for the allowed GS wave vector. The
GS with vanishing current may be reached by the method of twisted
boundary conditions
\cite{zotos}, resulting in the GS momentum displaced away from
$(\frac\pi2,\frac\pi2)$. The asymmetry of $n_{\bf k}$ found in small
clusters can thus be attributed to this displacement and is a finite
size effect.  In the thermodynamic limit in the system with
AFM order the GS momentum would coincide with $(\frac\pi2,\frac\pi2)$
and no asymmetry is expected in $n_{\bf k}$.

To get more insight into the structure of $\delta n_{\bf k}$, we
simplify the wave function, Eq.~(\ref{psi}), by keeping only the
one-magnon contributions.  The leading order contribution to $\delta
n_{\bf k}$ is
\begin{eqnarray}
\delta n_{\bf k}^{(1)}\!\!&=&\!-Z_{{\bf k}_0}M_{{\bf k}_0{\bf q}}
G_{{\bf k}_0}(\epsilon_{{\bf k}_0}\!\!-\!\omega_{\bf q})\bigl[2 u_{\bf q}\!+\!
M_{{\bf k}_0{\bf q}}
G_{{\bf k}_0}(\epsilon_{{\bf k}_0}\!\!-\!\omega_{\bf q})\bigr] \nonumber\\
&\sim& -8  Z_{{\bf k}_0}^2 J
{{\bf q}\cdot {\bf v} \over \omega^2_{\bf q}}
(1+ Z_{{\bf k}_0} {{\bf q}\cdot {\bf v}  \over
\omega_{\bf q}  }), \qquad {q\to0}, \label{dn1}
\end{eqnarray}
\noindent
with ${\bf q}={\bf k}-{\bf k}_0$ (or ${\bf k}-{\bf k}_0-{\bf Q}$) and
${\bf v}=t(\sin k_{0x},\sin k_{0y})$.  The momentum dependence of EMD,
contained in $\delta n_{\bf k}^{(1)}$, essentially captures well the full
numerical solution for the isotropic case, Fig.~2(b), as well as in the
Ising limit, Fig.~2(a). 
\noindent
\begin{figure}[htb]
\vskip -1.2 cm
\epsfig{file=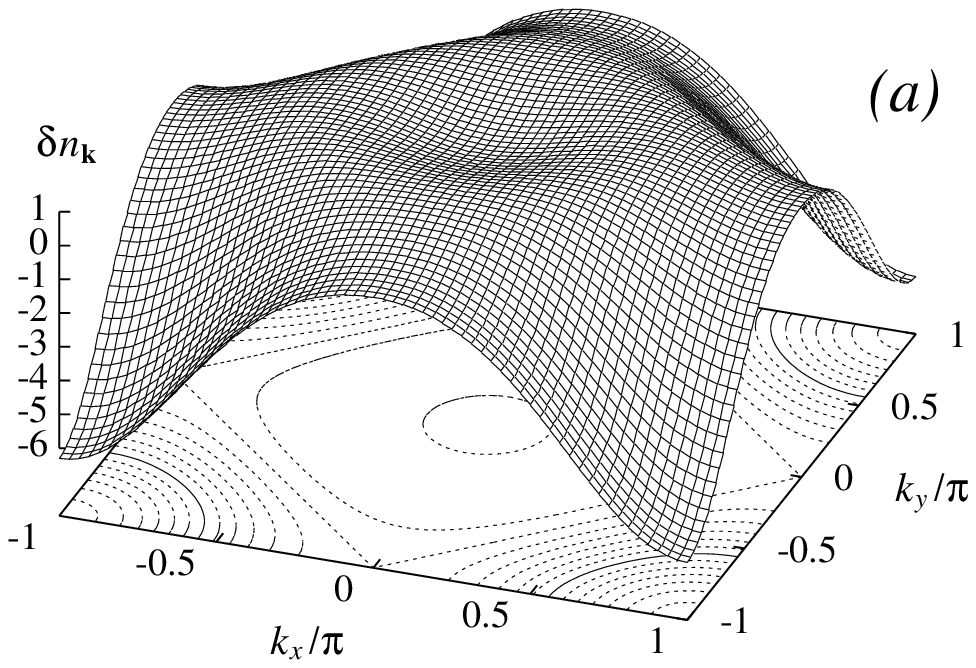,height=52mm,angle=0}
\vskip -1.4 cm 
\epsfig{file=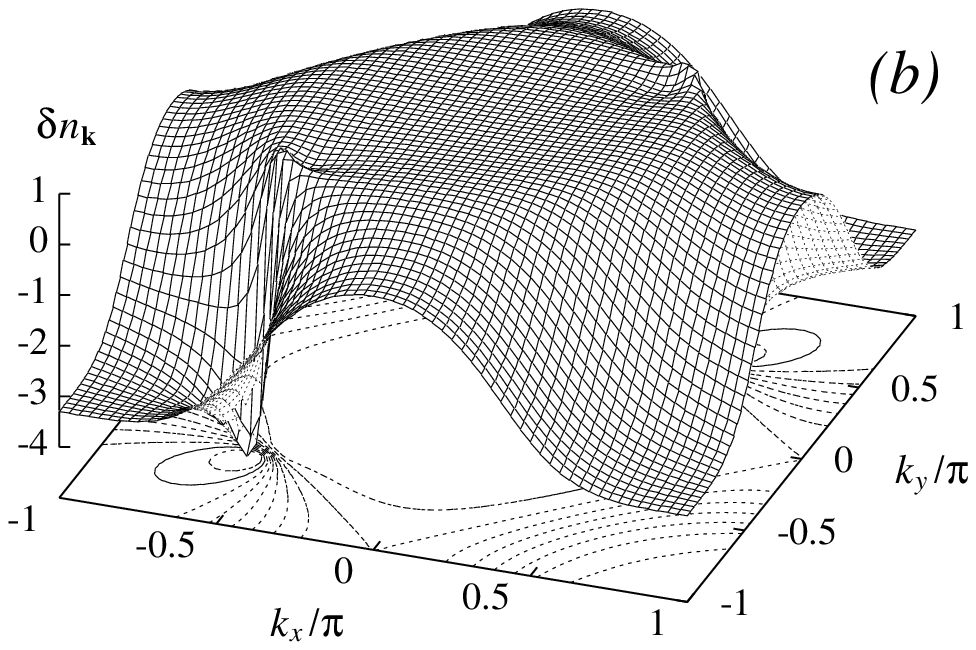,height=52mm,angle=0}
\vskip -1.4 cm 
\epsfig{file=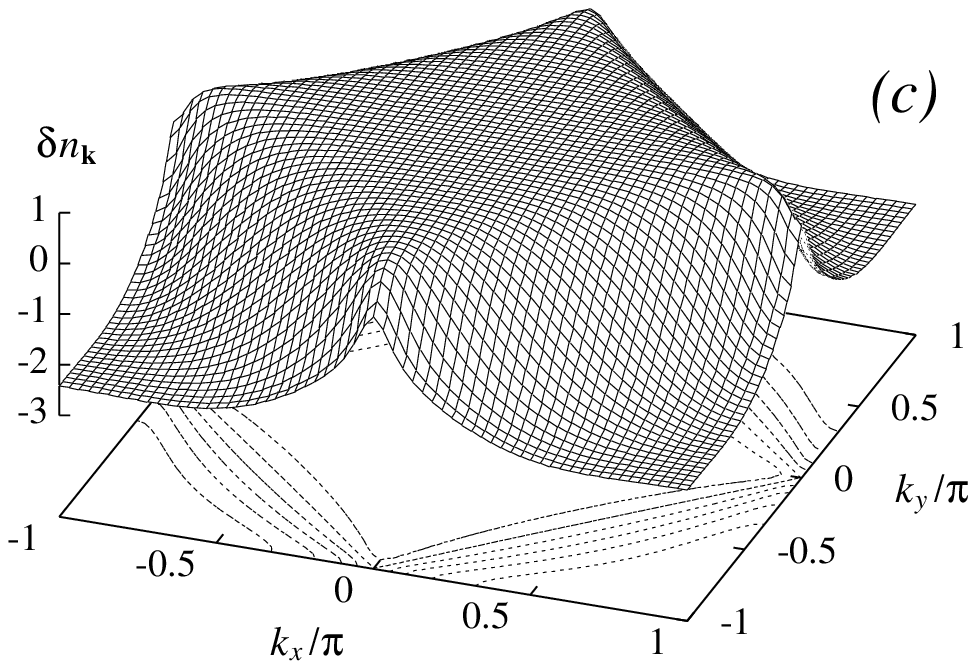,height=52mm,angle=0}
\caption{$\delta n_{\bf k}$ for (a) the Ising model $\gamma=0$, (b)
$\gamma=0.99$ and $t'\!=\!t''\!=\!0$,
and (c) $t$-$t'$-$J$ model with $t'=0.25 t$ and $\gamma=0.99$.  }
\end{figure}

A surprising observation is that the EMD exhibits in the extreme 
Heisenberg limit for momenta ${\bf k}\sim {\bf k}_0,
{\bf k}_0+{\bf Q}$ 
a discontinuity $\sim Z_{{\bf
k}_0} N^{1/2}$ and $\delta n_{\bf k}^{(1)} \propto -(1+{\rm sign}\,
q_x)/q_x$. These discontinuities are clearly seen in Fig.~1,
Fig.~2(b) and are consistent with ED results, Fig.~1. One
can interpret this result as an indication of an emerging {\it large}
Fermi surface at ${\bf k}\sim \pm {\bf k}_0$. 
The discontinuity appears only as {\it points} $\pm {\bf
k}_0$, not {\it lines} in the BZ. Note, however, that this
result is obtained in the extreme low doping limit, i.e., $c_{\rm
h}=1/N$ and it is not straightforward to generalize it to the finite doping
regime.   
In the limit $\gamma\to1$ this term does not strictly obey the
constraint $\delta n_{\bf k}
\leq 1$, although due to the symmetry it does not violate the EMD sum
rule.  The singularity is weak and on introducing a slight anisotropy,
e.g. $\gamma\alt 0.999$, the constraint is not violated.

In Fig.~3 we present the results for the
$t$-$t^\prime$-$t^{\prime\prime}$-$J$ model. First we
introduce {\it positive} next-nearest neighbor  
hopping matrix elements $t'=t/4$,
$t''=0$, claimed to be appropriate to 
electron doped systems such as Nd$_2$Ce
cuprates \cite{tohyama90}.  The GS is now twofold degenerate, with the
momenta at corners of
the AFM zone, e.g., ${\bf k}_0=(\pi,0)$, with an enhanced pole residue
$Z_{{\bf k}_0}=0.54$. The result is presented in Fig.~2(c) for
the entire BZ.  The discontinuity in this case disappears due to the
symmetry as evident from ${\bf v}=0$ in the leading order
approximation, Eq.~(\ref{dn1}). The effect of negative $t'=-t/4$, $t''=0$ is
relatively weak: the GS momentum remains at ${\bf
k}_0=(\frac\pi2,\frac\pi2)$, $\delta n_{\bf k}$ at ${\bf k}={\bf Q}$
is lower than the $t'=t''=0$ result while the discontinuity is
smaller, because $Z_{{\bf k}_0}=0.25$ here. In Fig.~3 we additionally
present results for $N_{\rm }(n_{\bf k}-1)$ obtained from exact
diagonalization of a $N_{\rm }=\sqrt{20}\times\sqrt{20}$ cluster.  The
$t'=t''=0$ results are in agreement with those of
Ref.~\cite{chernyshev98}. All ED results quantitatively confirm the
SCBA values. A possible set of parameters appropriate for reproducing
the dispersion from experimental ARPES data is $t'=-t/4$, $t''=t/5$
\cite{kyung96}.  Our SCBA result presented in Fig.~3 is qualitatively
similar to other $t'\leq 0$ results. The main difference is a more
pronounced step at ${\bf k}=\pm{\bf k}_0$.
\noindent
\begin{figure}[htb]
\vskip -3.4 cm \hskip 0 cm
\epsfig{file=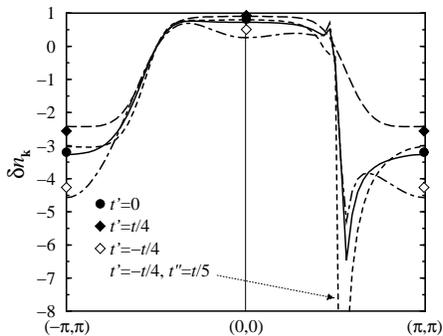,height=90mm,angle=-90}
\caption{$\delta n_{\bf k}$ for different $t^\prime$ and
$t^{\prime\prime}$. Symbols and lines represent ED for  $N_{\rm }=20$ sites
and SCBA results, respectively. Full circle and full line
denote $t'\!=\!t''\!=\!0$, full
diamonds and long dashed line $t'\!=\!t/4$, $t''\!=\!0$ and
empty diamonds, dot-dashed line
$t'\!=\!-t/4$, $t''\!=\!0$. Short dashed line represents the
SCBA result for $t'\!=\!-t/4$, $t''\!=\!t/5$. In all cases $\gamma=0.999$.}
\end{figure}
%

In the present work we considered the electron momentum distribution
function in underdoped cuprates. The results of the two methods, the
self consistent Born approximation and the exact diagonalization agree
quantitatively. Our analysis shows that the presence of next-nearest
neighbor terms changes EMD only quantitatively if the ground state is
at $(\frac\pi2,\frac\pi2)$ and qualitatively for sufficiently large
$t''>0$ where the GS momentum is at $(\pi,0)$.

The main observation is however the coexistence of two apparently
contradicting Fermi-surface scenarios in EMD of a single hole in an
AFM. (i) On one hand, the $\delta$-function contributions in Eq.(5) seem
to indicate that at finite doping a delta-function might develop into
small Fermi surface, i.e., a hole pocket, provided that AFM long range
order persists. (ii) A novel feature is that also $\delta n_{\bf k}$
is singular in a particular way, i.e., it shows a discontinuity at
${\bf k}={\bf k}_0$ with a strong asymmetry with respect to ${\bf
k}_0$. It is therefore more consistent with infinitesimally short arc
(point) of an emerging large FS. For finite doping the discontinuity could
possibly extend into such a finite arc (not closed) FS. Note that as
long-range AFM order is destroyed by doping, 'hole-pocket'
contributions should disappear while the singularity in $\delta n_{\bf
k}$ could persist.

Making contact with ARPES experiments we should note that ARPES
measures the imaginary part of the electron Green's function.  We must
note that using these experiments in underdoped cuprates $n_{\bf k}$
can be only qualitatively discussed since the latter is extracted only
from rather restricted frequency window below the chemical potential.
Nevertheless our results are not consistent with a small hole-pocket
FS (at least only a part of presumable closed FS is visible), but
rather with partially developed arcs resulting in FS which is just a
set of disconnected segments at low temperature collapsing to the
point \cite{norman98}. The SCBA results for singular $\delta n_{\bf
k}$ seem to allow for such a scenario. It should also be stressed that
the SCBA approach is based on the AFM long-range order, still we do
not expect that finite but longer-range AFM correlations would
entirely change our conclusions.
%

\end{document}